# A note on the solvable model of cosmology in the Earth-related coordinate system

Jian-Miin Liu
Department of Physics, Nanjing University
Nanjing, The People's Republic of China
On leave. E-mail: liu@phys.uri.edu

As a note for paper [physics/0505035], this is to make it more clear how we can depend upon Einstein's theory of gravitation and two assumptions of the cosmological principle and the perfect fluid, in the so-called Z-approximation, to construct a solvable model of cosmology in the Earth-related coordinate system. Hence, the cosmic acceleration becomes understandable.

The increasingly accurate cosmological measurements of the cosmic microwave background [1-4] and the high red-shift type Ia Supernovae [5-7] indicate an acceleration of the expansion of the Universe. This cosmic acceleration is unexpected in the standard model of cosmology [8-12]. It severely challenges the standard model of cosmology and urges cosmologists to make much theoretical corrections and attempts including to probe the nature of mysterious anti-gravitational dark energy or quintessence field [13-14], to modify Einstein's theory of gravitation [15-17] and to develop the inflation scenario [18-19]. As said, however, no well-developed and well-motivated solutions have yet emerged [15,16]. Actually, we attempted another way in our recent work [20-22]: to construct a solvable model of cosmology in the Earth-related coordinate system. This note is to make it more clear how we can depend upon Einstein's theory of gravitation and two assumptions of the cosmological principle and the perfect fluid, in the Earth-related (or Earth-based) coordinate system, to construct such a model.

From very beginning [23-26] and till now [8-12], to construct the standard model of cosmology, cosmologists adopted the co-moving coordinate system which, at any point in space, is at rest with respect to the matter located at that point. We adopted the Earth-related coordinate system, in which the Earth is always at rest, for the same purpose. The motivation to do so was brought about by the following considerations. (1) We are living on the Earth. Our cosmological measurements are all with respect to the Earth-related coordinate system. We should carry out our theoretical calculations in the Earth-related coordinate system to ensure a direct comparison between measurement data and calculation results. (2) The Earth-related coordinate system is at least approximately an inertial coordinate system if the discussed gravitational field of the Universe is turned off. That will provide with a useful constraint when we solve the Einstein field equation in the gravitational field of the Universe.

As is usual with the standard model of cosmology, we can start with two assumptions of the cosmological principle and the perfect fluid for the matter of the Universe. The cosmological principle states: With respect to the Earth-related frame of reference, the Universe is spatially homogeneous and isotropic on large scales. The assumption of the perfect fluid specifies the energy-momentum tensor for the matter of the Universe to be

$$T_\nu^\mu = (\rho + p)U^\mu U_\nu - p\delta_\nu^\mu, \quad \mu,\nu = 0,1,2,3, \qquad (1)$$

where $\rho$ and $p$ are respectively the fluid mass density and pressure measured by a co-moving observer, $U^\mu = dx^\mu/ds$, is the four-dimensional fluid velocity vector, $dx^0 = cdt$, $ds = \sqrt{g_{\mu\nu}(x)dx^\mu dx^\nu}$ is the differential space-time "distance", $c$ is the speed of light in vacuum. We take the system of units of $c=1$ unless specified.

Under the assumption of the perfect fluid, the energy-momentum tensor $T_\nu^\mu$ may depend on space-time coordinates $x^\mu, \mu = 0,1,2,3$, and fluid velocity, $u^r = dx^r/dt$, $r = 1,2,3$. But, the cosmological principle places more restrictions on this tensor. Tensor $T_\nu^\mu$ may depend only on time coordinate $t$ and fluid velocity magnitude $u = \sqrt{(u^1)^2 + (u^2)^2 + (u^3)^2}$ because of the large-scale homogeneity and isotropy in space. Moreover, this fluid velocity magnitude is the same everywhere in space again because of the large-scale spatial homogeneity. It must be the most probable velocity magnitude of the perfect fluid. We now take an approximation: the most probable velocity magnitude of the perfect fluid vanishes with respect to the Earth-related frame of reference. We call it the Z-approximation. The Z-approximation is reasonable since the most probable velocity magnitude of the perfect fluid is small as most (around 87 percent) of the gravity-interacting matter of the Universe is made up of dark matter that is cold and the cosmic background temperature is very low, 2.7K.

On the other hand, as shown in Ref.[20-22], the cosmological principle allows us to find the Earth-related coordinate system, denoted with $\{ct, r, \theta, \phi\}$, in which the line element for space-time of the Universe is determined up to two time-dependent factors,

$$ds^2 = S^2(t)c^2 dt^2 - R^2(t)\{\frac{dr^2}{1-kr^2} + r^2 d\theta^2 + r^2 \sin^2\theta d\phi^2\}, \qquad (2)$$
$$S^2(t) > 0, \ R^2(t) > 0,$$

where $S^2(t)$ and $R^2(t)$ are the two factors, $S^2(t) > 0$ and $R^2(t) > 0$ hold for keeping the space-time signature $(+,-,-,-)$, and $k$ is the spatial curvature which takes value of +1 or 0 or -1, separately characterizing a closed or flat or open space.

The cosmological principle also implies that the Universe is filled with the gravity-interacting matter which distributes uniformly in space, in other words, non-zero mass density ρ of the matter of the Universe is independent of spatial coordinates. It is independent of time coordinate, too. Otherwise, we will face the problem that mass would be created (or annihilated) with the same rate everywhere in an arbitrarily big but finite spatial region.

In the Earth-related coordinate system, owing to the Z-approximation, we have fluid velocity: $u^1 = dr/dt = 0$, $u^2 = d\theta/dt = 0$ and $u^3 = d\phi/dt = 0$. That gives arise to

$$U^0 = S^{-1}(t), \ U^r = 0, r = 1,2,3.$$

At the same time, $\rho$ and $p$ become the fluid mass density and pressure with respect to the Earth-related coordinate system. Consequently, the energy-momentum tensor for the matter of the Universe is simply

$$T_\nu^\mu = \begin{cases} \rho, & for \mu = \nu = 0, \\ -p\delta_\nu^\mu, & for \mu = \nu = 1,2,3, \\ 0, & otherwise, \end{cases} \qquad (3)$$

in the Earth-related coordinate system.

The Einstein field equation with a cosmological term is

$$R_\nu^\mu - \frac{1}{2} R_\lambda^\lambda \delta_\nu^\mu = 8\pi G T_\nu^\mu + \Lambda \delta_\nu^\mu, \quad \mu = \nu = 0,1,2,3, \qquad (4)$$

where G is the Newtonian gravitational constant, $\Lambda$ is the cosmological constant, $R_\nu^\mu$ is the Ricci curvature tensor and $R_\lambda^\lambda$ is the space-time curvature.

We can use Eq.(2) to compute the Ricci tensor and the space-time curvature. Putting them and Eq.(3) into Eq.(4), we obtain

$$\dot{R}^2 + k\,S^2 = \frac{8}{3}\pi\rho G\,R^2\,S^2 + \frac{1}{3}\Lambda\,R^2\,S^2,$$

$$2S\,R\,\ddot{R} + S\,\dot{R}^2 + k\,S^3 - 2R\,\dot{R}\,\dot{S} = -8\pi\,p\,G\,S^3\,R^2 + \Lambda\,S^3\,R^2,$$

as the dynamical equations or the Z-approximate Einstein field equation in the gravitational field of the Universe for our model of cosmology. This model is solvable receiving benefit from adopting the Earth-related coordinate system.

Even this solution enables us to understand the observed cosmic acceleration. Readers may refer to Ref.[20-22] for all the details.

**Acknowledgment**

The author greatly appreciates the teachings of Prof. Wo-Te Shen. The author thanks to Dr. J. Conway for his supports and helps.

**References**

[1]   A. T. Lee et al, Astrophys. J., 561, L1 (2001)
[2]   C. B. Netterfield et al, Astrophys. J., 571, 604 (2002) [astro-ph/0104460]
[3]   N. W. Halverson et al, Astrophys. J., 568, 38 (2002)
[4]   D. N. Speegel et al, astro-ph/0302209
[5]   A. G. Riess et al, Astron. J., 116, 1009 (1998) [astro-ph/9805201]
[6]   S. Perlmutter et al, Astrophys. J., 517, 656 (1999) [astro-ph/9812133]
[7]   J. L. Tonry et al, astro-ph/0305008
[8]   S. Dodelson, Modern Cosmology, Academic Press (New York, 2003)
[9]   S. K. Bose, An Introduction to General Relativity, Wiley & Sons (New York, 1980)
[10]  M. Trodden and S. M. Carroll, astro-ph/0401547
[11]  J. Lesgourgues, astro-ph/0409426
[12]  J. Garcia-Bellido, astro-ph/0502139
[13]  E. V. Linder, astro-ph/0406189
[14]  R. Scranton et al, astro-ph/0307335
[15]  S. M. Carroll et al, Phys. Rev., D70, 043528 (2004)
[16]  S. M. Carroll et al, astro-ph/0410031
[17]  M. Carmeli, astro-ph/0205396
[18]  E. W. Kolb et al, hep-th/0503117
[19]  A. V. Yurov and S. D. Vereshchagin, Theor. Math. Phys., 139, 787 (2004) [hep-th/0502099]
[20]  Jian-Miin Liu, physics/0505035
[21]  Jian-Miin Liu, physics/0505036
[22]  Jian-Miin Liu, Influences of the gravitational field of the Universe on propagation velocities of light signals coming from distant galaxies, to be published
[23]  A. Friedmann, Z. Phys., 10, 377 (1922)
[24]  G. Lemaitre, Ann. Soc. Sci. (Bruxelles), 47, 49 (1927)
[25]  H. P. Robertson, Appl. Phys. J., 82, 284 (1935); 83, 187 (1936); 83, 257 (1936)
[26]  A. G. Walker, Proc. Lond. Math. Soc., 42, 90 (1936)